\documentclass{iau}
\usepackage{graphicx}

\title[Gamma-ray emitting narrow-line Seyfert 1 galaxies]
{Gamma-ray emitting narrow-line Seyfert 1 galaxies. New discoveries and open questions}

\author[F. D'Ammando et al.]{F. D'Ammando$^{1}$\thanks{dammando@ira.inaf.it}, M. Orienti$^{1}$, J. Finke$^{2}$, J. Larsson$^{3}$, M. Giroletti$^{1}$, on behalf of the {\em Fermi} Large Area Telescope Collaboration} 

\affiliation{$^{1}$INAF - Istituto di Radioastronomia, Via Gobetti 101,
  I-40129 Bologna, Italy \\[\affilskip]$^{2}$U.S. Naval Research Laboratory, 4555 Overlook Ave. SW, Washington, DC 20375-5352, USA \\[\affilskip] $^{3}$KTH, Dep. of Physics, and the Oskar Klein Centre, AlbaNova, SE-106 91 Stockholm, Sweden}

\pubyear{2014}
\volume{304}
\pagerange{1}
\jname{Multiwavelength AGN Surveys and Studies}
\editors{Areg Mickaelian, Felix Aharonian and David Sanders}
\begin{document}

\maketitle

\begin{abstract}
The discovery of $\gamma$-ray emission from 5 radio-loud narrow-line Seyfert 1
galaxies revealed the presence of a possible emerging third class of AGNs with
relativistic jets, in addition to blazars and radio galaxies. The existence of
relativistic jets also in this subclass of Seyfert galaxies opened an
unexplored research space for our knowledge of the radio-loud AGNs. Here, we discuss the radio-to-$\gamma$-rays properties of the $\gamma$-ray emitting narrow-line Seyfert 1 galaxies, also in comparison with the blazar scenario.
\end{abstract}

\firstsection 

\section{Introduction}

Before the launch of the {\it Fermi} satellite only two classes of AGNs were known to produce relativistic jets and thus to emit
up to the $\gamma$-ray energy range: blazars and radio galaxies, both hosted
in giant elliptical galaxies \cite[(Blandford \& Rees 1978)]{blandford78}. The
first 5 years of observation by the Large Area Telescope (LAT) on board {\em
  Fermi} confirmed that the extragalactic $\gamma$-ray sky is dominated by
radio-loud AGNs, being mostly blazars and some radio galaxies. However, the
discovery by {\em Fermi}-LAT of variable $\gamma$-ray emission from a few radio-loud narrow-line Seyfert 1s (NLSy1s) revealed the presence of a possible
third class of AGNs with relativistic jets \cite[(Abdo et al.~2009)]{abdo2009c}. On the contrary, no radio-quiet
Seyfert galaxies were detected in $\gamma$ rays so far \cite[(Ackermann et al.~2012)]{ackermann12b}.

NLSy1 is a class of AGN identified by \cite[Osterbrock \& Pogge (1985)]{osterbrock85}
and characterized by their optical properties: narrow permitted lines (FWHM
(H$\beta$) $<$ 2000 km s$^{-1}$) emitted from the broad line region (BLR),
[OIII]/H$\beta$ $<$ 3, and a bump due to Fe II \cite[(see e.g.~Pogge 2000, for
a review)]{pogge00}. They also exhibit strong X-ray variability, steep X-ray
spectra, and substantial soft X-ray excess \cite[(e.g.~Grupe et al.~2010)]{grupe10}. These characteristics point to systems with smaller
masses of the central black hole (BH = 10$^6$--10$^8$ M$_\odot$) and higher accretion rates (close to or above the Eddington limit) with respect to blazars and radio
galaxies. NLSy1s are generally radio-quiet (radio-loudness $R<$ 10), with only
a small fraction of them \cite[($<$ 7$\%$; Komossa et al.~2006)]{komossa06} classified as radio-loud, and objects with high
values of radio-loudness ($R>$ 100) are even more sparse ($\sim$2.5\%), while
generally $\sim$15$\%$ of quasars are radio-loud. Considering also that NLSy1s
are thought to be hosted in spiral galaxies, their detection in $\gamma$ rays poses intriguing questions about the
nature of these sources, the production of relativistic jets, the
mechanisms of high-energy emission, and the cosmological evolution of radio-loud AGN.

\section{Radio-to-$\gamma$-ray properties of the $\gamma$-ray NLSy1s}

So far five radio-loud NLSy1 galaxies have been detected at high significance
by {\em Fermi}-LAT: 1H 0323$+$342, SBS 0846$+$513, PMN J0948$+$0022, PKS
1502$+$036, and PKS 2004$-$447 \cite[(Nolan et al.~2012, D'Ammando et
al.~2012)]{nolan12, dammando12}, with a redshift between 0.061 and 0.585. Their average apparent isotropic luminosity in the 0.1--100 GeV energy band is between
10$^{44}$ erg s$^{-1}$ and 10$^{47}$ erg s$^{-1}$, a range of values typical
of blazars \cite[(D'Ammando et al.~2013b)]{dammando13b}. This could be an
indication of a small viewing angle with respect to the jet axis and thus a high beaming
factor for the $\gamma$-ray emission, similarly to blazars. 
In particular, SBS 0846$+$513 and PMN J0948$+$0022 showed $\gamma$-ray flaring
activity combined with a moderate spectral evolution \cite[(D'Ammando et al.~2012, Foschini et al.~2011)]{dammando12,foschini11}, a behaviour
already observed in bright FSRQs and low-synchrotron-peaked BL Lacs
\cite[(Abdo et al.~2010)]{abdo10}. Several strong $\gamma$-ray flares were observed
from SBS 0846$+$513 and PMN J0948$+$0022, reaching at the peak an apparent isotropic $\gamma$-ray
luminosity of $\sim$10$^{48}$ erg s$^{-1}$, comparable to that of the bright
FSRQs \cite[(Foschini et al.~2011, D'Ammando et al.~2012, D'Ammando et al.~2013e)]{foschini11,dammando12,dammando13e}. Variability and spectral
properties of these two NLSy1s in $\gamma$ rays indicate a blazar-like
behaviour. Recently, an intense $\gamma$-ray flaring activity was observed by
LAT also from 1H 0323$+$342 \cite[(Carpenter et al.~2013)]{carpenter13}. This
is another indication that radio-loud NLSy1s are able to host relativistic jets as powerful as those in blazars.

Differently from the steep X-ray spectra usually observed in NLSy1s, a relatively hard X-ray spectrum was detected in the {\em Swift}-XRT
observations of SBS 0846$+$513 \cite[(D'Ammando et al.~2012, D'Ammando et al.~2013e)]{dammando12,dammando13e}, PMN J0948$+$0022 \cite[(Foschini et
al.~2011, D'Ammando et al.~2013d)]{foschini11, dammando13d}, 1H 0323$+$342 \cite[(D'Ammando et al.~2013c)]{dammando13c}, and PKS 1502$+$036
\cite[(D'Ammando et al.~2013a)]{dammando13a}. This suggests a significant contribution of inverse Compton radiation from a
relativistic jet, similar to what is found for FSRQs. The spectral modelling
of the XMM-{\em Newton} data of PMN J0948$+$0022 collected on 2011
May 28--29 showed that emission from the jet likely dominates the spectrum above $\sim$2~keV, while a soft X-ray excess is
evident in the low-energy part of the X-ray spectrum \cite[(D'Ammando et al.~2013d)]{dammando13d}.

When observed in the radio band with the high angular resolution provided by
the Very Long Baseline Array (VLBA), the $\gamma$-ray emitting NLSy1s usually show a
pc-scale core-jet structure. This is the case of SBS 0846$+$513 \cite[(D'Ammando et al.~2012)]{dammando12}, PKS
1502$+$036 \cite[(D'Ammando et al.~2013a)]{dammando13a}, and PMN J0948$+$0022
\cite[(Giroletti et al.~2011; D'Ammando et al.~2013d)]{giroletti11,dammando13d}, although in the last two sources the jet
structure is faint. The high variability brightness temperature estimated in
PKS 1502$+$03 \cite[($T_B$ = 2.5$\times$10$^{13}$ K; D'Ammando et al.~2013a)]{dammando13a}, in SBS 0846$+$513 \cite[($T_B$ =
1.1$\times$10$^{14}$ K; D'Ammando et al.~2013e)]{dammando13e}, and PMN
J0948$+$0022 \cite[($T_B$ = 3.4$\times$10$^{11}$ K; Giroletti et al.~2011)]{giroletti11}, strongly indicates that the jet has a Lorentz factor
larger than one. An indipendent proof on the presence of beaming effects is
the detection of superluminal motion of (9.3$\pm$0.6)$c$ in the jet of SBS
0846$+$513 \cite[(D'Ammando et al.~2013e)]{dammando13e}. On the contrary, VLBA observations did not detect apparent superluminal motion at 15 GHz for PKS
1502+036 during 2002--2012, although the radio spectral variability and the
one-sided structure seem to require the presence of boosting effects
in a relativistic jet \cite[(D'Ammando et al.~2013b)]{dammando13b}. 

A complex connection between the radio and $\gamma$-ray emission was observed
for SBS 0846$+$513 and PMN J0948$+$0022, where
$\gamma$-ray and radio flares have not a similar behaviour, as discussed in
detail in \cite[D'Ammando et al.~(2013e)]{dammando13e}, \cite[D'Ammando et al.~(2013d), and Foschini et al.~(2012)]{dammando13d,foschini12}.

The first spectral energy distributions (SEDs) collected for the four NLSy1s detected in the first year of {\em Fermi} operation showed
clear similarities with blazars: a double-humped shape with a first peak in the IR/optical band
due to synchrotron emission, a second peak in the MeV/GeV band likely due to inverse
Compton emission, and an accretion disc component in UV for three of the four sources. The physical
parameters of these NLSy1s are blazar-like, and the jet power is in the
average range of blazars \cite[(Abdo et al.~2009)]{abdo2009c}. 
The SEDs of two different activity states of SBS 0846$+$513,
modelled by an external Compton component of seed photons from a dust torus,
could be fitted by changing the electron distribution parameters as well as
the magnetic field \cite[(D'Ammando et al.~2013e)]{dammando13e}, consistent
with the modelling of different activity states of the blazar PKS\,0208$-$512
\cite[(Chatterjee et al.~2013)]{chatterjee13}. A significant shift of the
synchrotron peak to higher frequencies was observed in SBS 0846$+$513 during the 2012 May flaring episode, similar to
  FSRQs \cite[(e.g.~PKS\,1510$-$089; D'Ammando et al.~2011)]{dammando11}. Contrary to what is observed in PMN J0948$+$0022, no significant evidence of thermal emission from the accretion disc has been observed in SBS 0846$+$513 \cite[(D'Ammando et al.~2013e)]{dammando13e}.
 
\section{Radio-loudness, host galaxies, and jet formation}

The physical parameters that
drive the jet formation are still under debate. An important parameter could
be the BH mass, with only large masses allowing an
efficient jet formation \cite[(see e.g.~Sikora et al. 2007)]{sikora07}. Therefore one of the most
surprising fact related to the discovery of $\gamma$-ray emission from radio-loud NLSy1s
was the production of a relativistic jet in objects with a relatively small
BH mass \cite[(10$^{7}$--10$^{8}$ M$_{\odot}$; Yuan et al.~2008)]{yuan08}. However, it is worth noting that the mass
estimation of these sources has large
uncertainties. In particular, \cite[Marconi et al.~(2008)]{marconi08} suggested that BLR clouds are subjected
to radiation pressure from the absorption of ionizing photons. Applying a
correction for this effect on the virial BH masses, they obtained
higher values with respect to previous estimates for
the NLSy1s, which are objects radiating close to their Eddington limit. 
Recently, \cite[Calderone et al.~(2013)]{calderone13} modelling the optical/UV data of some
radio-loud NLSy1s with a Shakura \& Sunyaev disc spectrum estimated BH masses higher than 10$^{8}$ M$_{\odot}$. In particular, they derived a BH mass
of 10$^{9}$ M$_{\odot}$ and 2$\times$10$^{8}$ M$_{\odot}$ for PMN J0948$+$0022
and PKS 1502$+$036, respectively, in agreement with the typical BH mass of blazars. This
may solve the problem of the minimum BH mass predicted in different scenarios of
relativistic jet formation and development, but introduces a new issue. If the
BH mass of these NLSy1s is 10$^{8}$-10$^{9}$ M$_{\odot}$, how is it possible to
have such a large BH mass in a spiral galaxy?

Unfortunately only very sparse observations of the host galaxy of radio-loud NLSy1s are available up to
now. Among the NLSy1s detected by LAT only for the closest one, 1H
0323$+$342, the host galaxy was clearly detected. {\em Hubble Space Telescope}
and Nordic Optical Telescope observations seem to reveal a one-armed galaxy
morphology or a circumnuclear ring, respectively, suggesting two possibilities: the spiral arms of the host
galaxy \cite[(Zhou et al.~2007)]{zhou07} or the residual of a galaxy merging
\cite[(Anton et al.~2008)]{anton08}. 
Thus the possibility that the production of relativistic jets in these objects could be due to strong merger activity, unusual in disc/spiral galaxies, cannot be ruled out. 

According to the ``modified spin paradigm'', another fundamental
parameter for the efficiency of a relativistic jet production should be the BH
spin, with SMBHs in elliptical galaxies having on average much larger spins than SMBHs in spiral galaxies. This is due
to the fact that the spiral galaxies are characterized by multiple accretion
events with random angular momentum orientation and small increments
of mass, while elliptical galaxies undergo at least one major merger with
large matter accretion triggering an efficient spin-up of the SMBH. 
The accretion rate (thus the mass) and the spin of the BH seem to
be related to the host galaxy, leading to the hypothesis that relativistic jets
can form only in elliptical galaxy \cite[(e.g. Marscher et al.~2009, B\"ottcher et al.~2002)]{marscher09,bottcher02}. We
noted that the BH masses of radio-loud NLSy1s are generally larger than those in the entire sample of NLSy1s \cite[(M$_{\rm BH}$
$\approx$(2--10)$\times$10$^{7}$ M$_\odot$; Komossa et al.~2006, Yuan et al.~2008)]{komossa06,yuan08}, even if still small if compared to radio-loud quasars. The larger BH masses of radio-loud NLSy1s with
respect to radio-quiet NLSy1s could be related to prolonged accretion episodes that can spin-up the BHs. In
this context, the small fraction of radio-loud NLSy1s with respect to radio-loud quasars could be
an indication that not in all of the former the high-accretion regime lasted
long enough to spin-up the central BH \cite[(Sikora 2009)]{sikora09}.

\begin{acknowledgement}
The {\em Fermi} LAT Collaboration acknowledges support from a number of agencies and institutes for
both the development and the operation of the LAT as well as
scientific data analysis.  These include NASA and DOE in the United
States, CEA/Irfu and IN2P3/CNRS in France, ASI and INFN in
Italy, MEXT, KEK, and JAXA in Japan, and the
K.~A.~Wallenberg Foundation, the Swedish Research Council and the National Space Board in Sweden. Additional support from INAF in Italy
and CNES in France for science
analysis during the operations phase is also gratefully acknowledged. 
FD, MO, MG acknowledge financial contribution from grant PRIN-INAF-2011. 
\end{acknowledgement}


\begin{thebibliography}{} 

\bibitem[Abdo et al.(2009c)]{abdo2009c} Abdo, A. A., Ackermann, M., Ajello, M., et al. 2009, \textit{ApJ}, 707, L142
\bibitem[Abdo et al.(2010)]{abdo10} Abdo, A. A., Ackermann, M., Ajello, M., et al. 2010, \textit{ApJ}, 710, 1271
\bibitem[Ackermann et al.(2012)]{ackermann12} Ackermann, M., Ajello, M., Allafort, A., et al. 2012, \textit{ApJ}, 747, 104
\bibitem[Anton et al.(2008)]{anton08} Anton, S., Browne, I. W. A., Marcha, M. J. 2008, \textit{A\&A}, 490, 583
\bibitem[B{\"o}ttcher \& Dermer(2002)]{bottcher02} B{\"o}ttcher, M., \& Dermer, C.~D. 2002, \textit{ApJ}, 564, 86
\bibitem[Blandford \& Rees(1978)]{blandford78} Blandford, R.~D., \& Rees, M.~J. 1978, in \textit{BL Lac Objects ed. A.~M. Wolfe}, 328
\bibitem[Calderone et al.(2013)]{calderone13} Calderone, G., Ghisellini, G., Colpi, M., Dotti, M. 2013, \textit{MNRAS}, 431, 210
\bibitem[Carpenter et al.(2013)]{carpenter13} Carpenter, B., \& Ojha, R. 2013, \textit{The Astronomer's Telegram}, 5344, 1
\bibitem[Chatterjee et al.(2013)]{chatterjee13} Chatterjee, R., Fossati, G., Urry, C. M., et al. 2013, \textit{ApJ}, 763, L11
\bibitem[D'Ammando et al.(2011)]{dammando11} D'Ammando, F., Raiteri, C. M., Villata, M., et al. 2011, \textit{A\&A}, 529, 145
\bibitem[D'Ammando et al.(2012)]{dammando12} D'Ammando, F., Orienti, M., Finke, J., et al. 2012, \textit{MNRAS}, 426, 317
\bibitem[D'Ammando \& Orienti(2013)]{dammando13a} D'Ammando, F., Orienti, M., Doi, A., et al. 2013a, \textit{MNRAS}, 433, 952
\bibitem[D'Ammando \& Orienti(2013)]{dammando13b} D'Ammando, F., Tosti, G., Orienti, M., Finke, J. 2013b, \textit{2012 Fermi Symposium proceedings}
\bibitem[D'Ammando et al.(2013)]{dammando13c} D'Ammando, F., Carpenter, B., Ojha, R. 2013c, \textit{The Astronomer's Telegram}, 5352, 1
\bibitem[D'Ammando et al.(2013)]{dammando13d} D'Ammando, F., Larsson, J., Orienti, M., et al. 2013d, submitted to \textit{MNRAS}
\bibitem[D'Ammando et al.(2013)]{dammando13e} D'Ammando, F., Orienti, M., Finke, J., et al. 2013e, \textit{MNRAS}, 436, 191 
\bibitem[Foschini et al.(2011)]{foschini11} Foschini, L., Ghisellini, G., Kovalev, Y. Y., et al. 2011, \textit{MNRAS}, 413, 1671
\bibitem[Foschini et al.(2012)]{foschini12} Foschini, L., Angelakis, E., Fuhrmann, L., et al. 2012, \textit{A\&A}, 548, A106
\bibitem[Giroletti et al.(2011)]{giroletti11} Giroletti, M., Paragi, Z., Bignall, H., et al. 2011, \textit{A\&A}, 528, L11
\bibitem[Grupe et al.(2010)]{grupe10} Grupe, D., Komossa, S., Leighly, K. M., Page, K. L. 2010, \textit{ApJS}, 187, 64
\bibitem[Komossa et al.(2006)]{komossa06} Komossa, S., Voges, W., Xu, D., et al. 2006, \textit{AJ}, 132, 531
\bibitem[Marconi et al.(2008)]{marconi08} Marconi, A., Axon, D. J., Maiolino, R., et al. 2008, \textit{ApJ}, 678, 693
\bibitem[Marscher(2009)]{marscher09} Marscher, A. 2009, in \textit{Lecture Notes in Physics 794, ed. T. Belloni}, 173
\bibitem[Nolan et al.(2012)]{nolan12} Nolan, P., Abdo, A. A., Ackermann, M., et al. 2012, \textit{ApJS}, 199, 31 
\bibitem[Osterbrock \& Pogge(1985)]{osterbrock85} Osterbrock, D. E., \& Pogge, R. W. 1985, \textit{ApJ}, 297, 166  
\bibitem[Pogge(2000)]{pogge00} Pogge, R. W. 2000, \textit{New Astron. Revs}, 44, 381
\bibitem[Sikora et al.(2007)]{sikora07} Sikora, M., Stawarz, L., Lasota, J.-P. 2007, \textit{ApJ}, 658, 815
\bibitem[Sikora(2009)]{sikora09} Sikora, M. 2009, \textit{AN}, 330, 291
\bibitem[Yuan et al.(2008)]{yuan08} Yuan, W., Zhou, H.-Y., Komossa, S., et al. 2008, \textit{ApJ}, 685, 801
\bibitem[Zhou et al.(2006)]{zhou06} Zhou, H.-Y., Wang, T.-G., Yuan, W., et al. 2006, \textit{ApJS}, 166, 128
\end{thebibliography}
\end{document}